\documentstyle[12pt]{article}
\begin{document}

\bigskip
\begin{center}
Multiplicity distributions of quark and gluon jets in QCD\\
\bigskip
I.M. Dremin$^1$, J. Tran Thanh Van$^2$\\
\vspace{3mm}
{\it $^1$Lebedev Physical Institute, Moscow 117924, Russia\\
$^2$LPTHE, Univ. Paris-Sud F-91405 Orsay, Cedex, France \\}

\bigskip
\end{center}
\begin{abstract}
The solution of the equations for the generating functions of 
multiplicity 
distributions of quark and gluon jets in higher order 3NLO 
perturbative QCD is
obtained. The non-perturbative effects are discussed as well. The 
results
are compared to recent experimental data on mean multiplicities, 
their ratio
and its slope, higher moments of the distributions.
\end{abstract}

Here we briefly describe recent advances in theoretical 
understanding of
multiplicity distributions of quark and gluon jets. The extended 
survey 
with more detailed comparison to experimental data can
be found in the recent review paper \cite{dg00}.

The progress in experimental studies of properties of quark and 
gluon jets is
very impressive. Therefore the study of
the energy evolution of such parameters of multiplicity distributions 
of jets
as their average multiplicities and widths becomes possible. It is 
well known
that the average multiplicities of quark and gluon jets increase quite 
fast
with energy but their ratio has a much slower dependence. 

The perturbative QCD provides quite definite
predictions which can be confronted to experiment. In brief, the 
results can be 
summarized by saying that the energy dependence of the mean jet
multiplicity can be perfectly fitted but the ratio of gluon to quark jet
multiplicities can be described within the precision of 15-20$\%$ 
only.
Moreover, one can understand why next-to-leading approximation 
is good enough
for describing the energy dependence, but it is not quite 
satisfactory yet for
the ratio value. We show this by presenting the analytical 
expressions. For the
corresponding Figures, we refer the reader to the review paper 
\cite{dg00}.

The theoretical asymptotical value of the ratio of
average multiplicities equal 2.25 is much higher than its
experimental values, which are in the range from 1.05 at 
comparatively low energies of $\Upsilon $ resonance to 1.5 at 
$Z^0$ resonance. 
The next-to leading order (NLO) corrections reduce this ratio from 
its
asymptotical value by about 10$\%$ at $Z^0$ energy. The NNLO 
and 3NLO
terms diminish it further and show the tendency to approximate the 
data 
with better accuracy.
The computer solution of QCD equations for the generating 
functions has shown 
even better agreement with experiment not only on this ratio but on
higher moments of multiplicity distributions as well. Being perfect
at $Z^0$ energy, the agreement in the ratio is not as good at lower 
energies
where the theoretical curve is still about 15-20$\%$
above the experimental one. In other words, the theoretically 
predicted 
{\em slope of the ratio} of multiplicities in gluon and quark jets is
noticeably smaller than its experimental value. Nevertheless, one 
can speak 
about the steady convergence of theory and experiment with 
subsequent 
improvements being done. Moreover, it is even surprising that any 
agreement
is achieved in view of the expansion parameter being extremely 
large
(about 0.5) at present energies.

The importance of studying the slopes stems from the fact that 
some of them 
are extremely sensitive (while others are not) to higher order
perturbative corrections and to non-perturbative terms in the 
available energy
region. Thus they provide us with a good chance to learn more
about the structure of the perturbation series from experiment. 

In the perturbative QCD, the general approach to studying the 
multiplicity
distributions is formulated in the framework of equations for 
generating
functions. Therefrom, one can get equations
for average multiplicities and, in general, for any moment of the 
multiplicity
distributions \cite{dr3, dr4}. In particular, two equations for average
multiplicities of gluon and quark jets are written as
\begin{eqnarray}
\langle n_G(y)\rangle ^{'} =\int dx\gamma _{0}^{2}[K_{G}^{G}(x)
(\langle n_G(y+\ln x)\rangle +\langle n_G(y+\ln (1-x)\rangle -\langle 
n_G(y)
\rangle ) \nonumber  \\
+n_{f}K_{G}^{F}(x)(\langle n_F(y+\ln x)\rangle +\langle n_F(y+
\ln (1-x)\rangle -\langle n_G(y)\rangle )],  \label{ng}
\end{eqnarray}
\begin{equation}
\langle n_F(y)\rangle ^{'} =\int dx\gamma _{0}^{2}K_{F}^{G}(x)
(\langle n_G(y+\ln x)\rangle +\langle n_F(y+\ln (1-x)\rangle -\langle 
n_F(y)
\rangle ).   \label{nq}
\end{equation}
Herefrom one can learn about the energy evolution
of the ratio of multiplicities $r$ and of the QCD anomalous 
dimension $\gamma $
(the slope of the logarithm of average multiplicity in a gluon jet) 
defined as
\begin{equation}
r=\frac {\langle n_G\rangle }{\langle n_F\rangle }\; ,\;\;\;\;\; \;\;\;
\gamma =\frac {\langle n_G\rangle ^{'}}{\langle n_G\rangle }
=(\ln \langle n_G\rangle )^{'}\; .  \label{def}
\end{equation}
Here, prime denotes the derivative over the evolution parameter
$y=\ln (p\Theta /Q_{0}),\\ 
p, \,\Theta $ are the momentum and the initial angular
spread of the jet, related to the parton virtuality $Q=p\Theta /2$,
\, $Q_{0}$=const, \, $K$'s are the well known splitting
functions, $\langle n_G\rangle $ and
$\langle n_F\rangle $ are the average multiplicities in gluon and 
quark jets, \,
$\langle n_G\rangle ^{'}$ is the slope of $\langle n_G\rangle $, \, 
$n_f$ is
the number of active flavours. The perturbative expansion of 
$\gamma $ and 
$r$ is written as
\begin{equation}
\gamma =\gamma _{0}(1-a_{1}\gamma _0 -a_{2}\gamma _{0}^{2} -
a_{3}\gamma _{0}^{3}
)+O(\gamma _{0}^{5}),  \label{gam}
\end{equation}
\begin{equation}
r=r_0(1-r_{1}\gamma _{0}-r_{2}\gamma _{0}^{2}-r_{3}\gamma 
_{0}^{3})+O(\gamma 
_{0}^{4}),   \label{rat}
\end{equation}
where $\gamma _{0}=\sqrt {2N_{c}\alpha _{S}/\pi }, \, \alpha _{S}$
is the strong coupling constant, 
\begin{equation}
\alpha_{S}=\frac {2\pi }{\beta _{0}y}\left [1-\frac {\beta _{1}\ln (2y)}
{\beta _{0}^{2}y}\right ]+O(y^{-3}),   \label{alp}
\end{equation}
$\beta _{0}=(11N_{c}-2n_{f})/3, \, \beta _{1}=(51N_{c}-19 n_{f})/3, 
r_0 = N_c/C_F,$ \,and in QCD $N_{c}=3$ is the number of colours, 
$C_{F}=4/3$. 

The limits of integration in eqs. (\ref{ng}), (\ref{nq}) used to be 
chosen 
equal either to 0 and 1 or to $e^{-y}$ and $1-e^{-y}$. 
This difference, being negligibly small at high energies $y$, is quite
important at low energies. Moreover, it is of physics significance. 
With limits
equal to $e^{-y}$ and $1-e^{-y}$, the partonic cascade terminates 
at the 
perturbative level $Q_0$ as is seen from the arguments of 
multiplicities in the 
integrals. With limits equal to 0 and 1, one extends the cascade 
into the 
non-perturbative region with low virtualities $Q_{1}\approx xp\Theta 
/2$ and
$Q_{2}\approx (1-x)p\Theta /2$ less than $Q_{0}/2$. This region 
contributes
terms of the order of $e^{-y}$, power-suppressed in energy. It is not 
clear 
whether the equations and LPHD hypothesis are valid down to 
some $Q_0$
only or the non-perturbative region can be included as well.

Nevertheless, the purely perturbative expansion (\ref{gam}), (\ref{rat})
with constant coefficients $a_{i}, r_{i}$ and energy-dependent 
$\gamma _0$
is at work just in the case of limits 0 and 1. 
The values of $a_i , r_i$ for different number of
active flavors $n_f$ are tabulated in \cite{cdgnt}. 
At $Z^0$-energy the subsequent terms in (\ref{rat})
diminish the value of $r$ compared with its asymptotics $r_0 
=2.25$
approximately by $10\%,\, 13\%,\, 1\%$ for $n_f=4$ getting closer 
to
experiment. However the theoretical
value of $r$ still exceeds its experimental values by 15-20$\%$. At 
lower
energies the value of $r$ diminishes due to the running property of 
the
coupling strength. Thus these calculations are more sensitive to it 
than
the fractality studies where it is almost unnoticed at large bins 
(e.g., see \cite{ddki, ddre}).

The energy dependence of mean multiplicities can be obtained 
\cite{dgpl, cdgnt}
from the definition (\ref{def}) by inserting there the value of $\gamma 
$
(\ref{gam}) and integrating over $y$. Keeping the terms as small as 
$y^{-1}$
at large $y$ in the exponent, one gets \cite{cdgnt} the following 
expressions
for energy dependence of multiplicities of gluon (G) and quark (F) 
jets
\begin{equation}
\langle n_G\rangle =Ky^{-a_1C^2}\exp 
   \left[ 2C\sqrt y +\delta _G(y) \right], \label{ngy}
\end{equation}
with $K$ an overall normalization constant, $C=\sqrt {4N_c/\beta 
_0}$, and
\begin{equation}
\delta _G(y)=\frac {C}{\sqrt y}\left [ 2a_2C^2+\frac {\beta _1}{\beta 
_0^2}
[\ln (2y)+2]\right ] 
+\frac {C^2}{y}\left [ a_3C^2-\frac {a_1\beta _1}{\beta _0^2}
[\ln (2y)+1]\right ];   \label{del}
\end{equation}
\begin{equation}
  \langle n_F\rangle =\frac {K}{r_0}y^{-a_1C^2}\exp 
     \left[ 2C\sqrt y +\delta _F(y)\right], 
\label{nfy}
\end{equation}
with
\begin{equation}
\delta _F(y)=\delta _G(y)+\frac {C}{\sqrt y}r_1+\frac 
  {C^2}{y}(r_2+\frac {r_1^2}
  {2}).   \label{dfy}
\end{equation}
It happens that 2NLO and 3NLO terms (contributing to $y^{-1/2}$ 
and $y^{-1}$ 
terms in the exponent)
are almost constant at present energies and do not change the 
energy dependence
prescribed in NLO approximation. It explains why the energy 
dependence is well 
fitted by both NLO and 3NLO formulas while  2NLO correction to 
the ratio $r$ is large and
important. 

The rather small difference in $r$ values results
in quite noticeable disagreement of the slopes $r'$. 
Theoretical estimates can be shown \cite{cdgnt} to be quite 
predictive for 
the ratio of the slopes of multiplicities but it is much less
reliable to use the perturbative estimates even at $Z^0$-energy for 
such
quantities as the slope of $r$ or the ratio of slopes of logarithms of 
multiplicities (the logarithmic slopes). Much higher 
energies are needed to do that. Thus the values of $r^{'}$ and/or of
the logarithmic slopes can be used
to verify the structure of the perturbative expansion.

We demonstrate it here on the example of the slope value.
The slope $r'$ is extremely sensitive to higher 
order perturbative corrections. The role of higher order corrections 
is 
increased here compared with $r$ because each $n$th order term 
proportional
to $\gamma _{0}^{n}$ gets an additional factor $n$ in front of it when
differentiated, the main constant term disappears and the large ratio
$r_2/r_1$ becomes crucial:
\begin{equation}
r^{'} =Br_{0}r_{1}\gamma _{0}^{3}\left [1+\frac {2r_{2}\gamma _{0}}
{r_1}+\left (\frac {3r_3}{r_1}+B_{1}\right )\gamma 
_{0}^{2}+O(\gamma _{0}^{3})
\right ],       \label{rpri}
\end{equation}
where  the relation $\gamma _{0}^{'}\approx -B\gamma 
_{0}^{3}(1+B_{1}\gamma 
_{0}^{2})$ has been used with $B=\beta _{0}/8N_c ;\, 
B_{1}=\beta _{1}/4N_c\beta _0$. The factor in front of the bracket is 
very small
already at present energies: $Br_0r_1\approx 0.156$ and $\gamma 
_0\approx 0.5$.
However, the numerical estimate of $r^{'}$ is still indefinite due to 
the 
expression inside the brackets.
Let us note that each differentiation leads to a factor $\alpha _S$ or
$\gamma _{0}^{2}$, i.e., to terms of higher order.
For values of $r_1$, $r_2$, $r_3$ tabulated above ($n_f=4$) one 
estimates
$2r_{2}/r_{1}\approx 4.9$, $(3r_3/r_1)+B_1\approx 1.5$).
The simplest correction proportional to $\gamma _0$ is more than 
twice larger 1 
at energies studied and the next one is about 0.4. Therefore
the ever higher order terms should be calculated for the perturbative 
values
of $r'$ to be trusted. 
The slope $r'$ is equal to 0 for a fixed coupling constant.

The higher order terms are also important for the moments of the 
multiplicity distributions \cite{dln}.
The normalized second factorial moment $F_2$ defines the width 
of the
multiplicity distribution. 

The asymptotical $(\gamma _{0}\rightarrow 0)$ 
values of $F_{2}^{G}$ and $F_{2}^{F}$ are different:
\begin{equation}
F_{2, as}^{G}=\frac {4}{3}, \;\;\;\; F_{2, as}^{F}=\frac {7}{4}.  
\label{fas}
\end{equation}
At $Z^0$ energy the widths of the distributions are smaller
\begin{equation}
F_{2}^{G}\approx 1.12, \;\;\;\; F_{2}^{F}\approx 1.34.   \label{fnum}
\end{equation}
but still much larger than their experimental values 1.02 and 1.08,
correspondingly. 
The rather large difference of the perturbative (\ref{fnum}) and 
experimental 
values at $Z^0$ indicates that moments of the distributions should 
be
sensitive to corrections. 
The conclusions about the third moments are similar. Nonetheless, 
the computer
solution of the QCD equation \cite{loch, olup} happened to be quite
successful in fitting experimental data even for higher moments and 
their 
ratios $H_q$ introduced in \cite{dr1}. It shows that the role of 
conservation 
laws treated approximately in the analytical approach and 
accurately accounted 
in computer calculations becomes more important for higher 
moments.

Thus it is shown that the analytical perturbative approach is quite 
successful
in demonstrating that all features of QCD predictions about 
multiplicities of
quark and gluon jets correspond to the general trends of 
experimental data.
Some disagreement at the level of 15-20$\%$ is understandable 
due to 
incomplete account for the energy-momentum conservation in such 
an approach.
Further accurate computer solutions are needed to check if these 
trends
persist at the higher precision level. They would take into account 
some
non-perturbative effects as well. The dipole model of QCD which 
includes 
implicitly the non-perturbative string effects has been developed
\cite{egus, egkh} in various versions. It describes experimental data 
even
with higher precision at present energies \cite{hkls}.

\end{document}